\documentclass{article}
\usepackage{graphicx} % Required for inserting images
\usepackage[HTML]{xcolor} 
\usepackage{amsmath}
\usepackage{mathabx}
\usepackage{geometry}

\usepackage{authblk} % Optional for a cleaner affiliation formatting
\usepackage[normalem]{ulem}
\usepackage{lineno}
\usepackage{placeins}

\begin{document}
\title{Spatio-temporal activity patterns induced by triadic interactions in an {\em in silico} neural medium}
\author[1,*]{Ana P. Millán}
\author[2]{Hanlin Sun}
\author[1,+]{Joaquín J. Torres}

\affil[1]{Institute ``Carlos I'' for Theoretical and Computational Physics, and Department of Electromagnetism and Physics of the Matter,
University of Granada, E-18071 Granada, Spain}
\affil[2]{Nordita, KTH Royal Institute of Technology and Stockholm University, Hannes Alfvéns väg 12, SE-106 91 Stockholm, Sweden}
\affil[*]{email:apmillan@ugr.es}
\affil[+]{email:jtorres@onsager.ugr.es}

\date{}

\maketitle
\begin{abstract}

Triadic interactions are general mechanisms by which a node or neuron can regulate directly the link or synapse between other two neurons. The regulation takes place in a familiar way by either depressing or facilitating synaptic transmission. 
Such interactions are ubiquitous in neural systems, accounting for axo-axonic synapses and tripartite synapses mediated by astrocytes, for instance, and have been related to neuronal and synaptic processes at different time-scales, including short and long-term synaptic plasticity. 
In the field of network science, triadic interactions have been shown to produce complex spatio-temporal patterns of connectivity. 
Here, we investigate the emergent behavior of an {\em in silico} neural medium constituted by a population of leaky integrate-and-fire neurons with triadic interactions. 
We observe that, depending on relevant parameters defining triadic interactions, different activity patterns emerge. 
These include i) a silent phase, 
ii) a low-activity phase in which complex spatio-temporal patterns of low neuronal firing rate emerge that propagate through the medium, 
iii) a high-activity phase characterized by complex spatio-temporal patterns of high neuronal firing rate that propagate through the neural medium as waves of high firing activity over a bulk of low activity neurons, 
and iv) a pseudo-blinking phase in which the neural medium switches between high and low activity states, in a similar fashion to up/down state transitions. 
Here we analyse in depth the features of such patterns and relate our findings to the recently proposed model of triadic percolation. 
    
\end{abstract}

%\linenumbers
\section{Introduction}

%Intro general

In recent decades, great experimental, theoretical, and computational effort has been devoted to understanding brain function, from elucidating the mechanisms underlying high-level brain functions to understanding the origin of brain disorders. 
Extensive literature investigates the biophysical processes that occur at the neuron, synapse and network developing levels, including the emergence of neuronal oscillations \cite{buzsaki2004neuronal} and avalanches \cite{beggs2003neuronal}, as well as synaptic plasticity at different time scales, namely short- \cite{zucker2002short} and long-term \cite{tsodyks1997neural} plasticity, and the underlying mechanisms responsible for synaptic pruning and brain development \cite{Johnson2010,millan2018concurrence,ana2021}. 
As a result of this research, mathematical models of neural networks have been developed to closely mimic neuronal populations \cite{DAngelo2013}. 
These models reproduce behavior observed in \textit{in vitro} and \textit{in vivo} experiments and can be analyzed using tools from statistical physics, dynamical systems, and Information Theory \cite{gused1,gused2}. 
Despite these advances, how other biophysical mechanisms influence complex emergent behavior in the brain, particularly in relation to higher-level cognitive functions, remains poorly understood. 

% Brain networks are dynamics
The brain, with its inherent complexity, is a prominent example of a dynamic networked system, where activity evolves over time giving rise complex spatio-temporal patterns \cite{Song2018,Townsend2018,Meyer-Baese2022}. 
Crucially, neuronal networks are not static but change both structurally and functionally through synaptic plasticity: the ability of synapses to strengthen or weaken over time in response to neuronal activity \cite{kandelbook}.
Synaptic plasticity underlies the brain's adaptability and is a critical mechanism for learning and memory. It occurs over varying time scales, broadly classified into short-term and long-term plasticity. Short-term synaptic plasticity involves rapid and transient changes in synaptic strength, lasting from milliseconds to minutes, and can both increase (facilitation) and decrease (depression) synaptic transmission in cases of repeated activation of the synapse, depending on the frequency of a presynaptic stimulus \cite{zucker2002short, tsodyks1997neural}. Short-term plasticity plays a key role in filtering information flow during high-frequency neuronal firing and it has been demonstrated to have important computational implications ~\cite{torres2013}. 
In contrast, long-term synaptic plasticity involves enduring changes in synaptic efficacy. Long-term potentiation refers to the strengthening of synapses following repeated activation (accounting for the Hebbian learning rule ``neurons that fire together wire together'' \cite{caporale2008spike}), and is often considered the cellular basis for learning and memory. Conversely, long-term depression refers to the weakening and ultimately pruning (or removal) of inactive synapses \cite{kandelbook}. 
Beyond synaptic efficacy, synaptic plasticity also tailors the brain's structural network. One key example during development is synaptic pruning, by which weak synapses are eliminated. This structural change is heavily modulated by astrocytes, a type of glial cell, which regulate synaptic transmission and plasticity through the release of gliotransmitters \cite{Chung2015,astro2021,Lee2021}.  
Such activity-driven mechanisms create feedback loops between brain architecture and activity that refine network performance over time \cite{millan2018concurrence, millan2019memory, millan2021growth}.  

\begin{figure}[t!]
    \centering
    \includegraphics[width=0.95\textwidth]{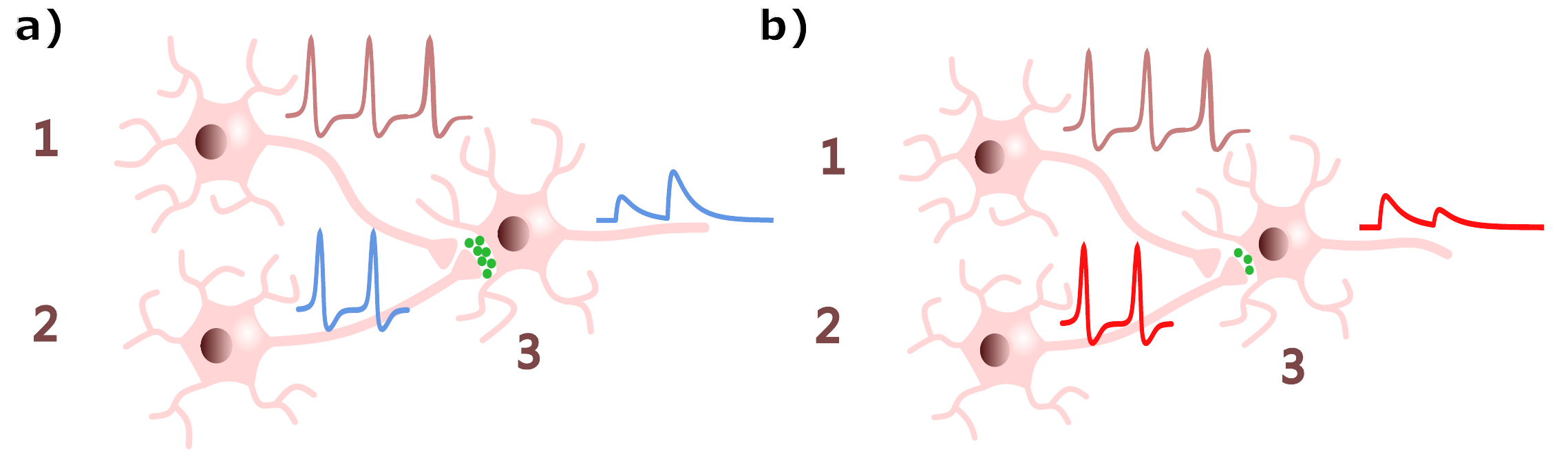}
    \caption{Cartoon illustrating particular triadic interactions among neurons. We illustrate the mechanism of axo-axonic facilitation (left) and inhibition (right). The former enhances the postsynaptic response in neuron $3$ due to the effect of the spiking activity of neuron $1$ arriving to the synapse $(2,3)$. 
    The later decreases the postsynaptic response in $3$ due the the negative regulation of synapse $(2,3)$ by the firing activity of neuron $1$.}
    \label{fig:Triadic_interactions}
\end{figure}

%Plasticity is triadic 
In general, synaptic plasticity mechanisms describe changes on synaptic efficacy due to the activity of the pre- and post-synaptic neurons. 
However, increasing evidence suggests that \emph{triadic interactions} also play a significant role in neuronal networks functioning \cite{Cho2016-br}. 
In such interactions, a third node (neuron) modulates the link (synapse) between two other nodes, generalizing the mechanisms of synaptic plasticity. 
Prominent examples are axo-axonic interactions (see figure \ref{fig:Triadic_interactions})\cite{kandelbook}, where the axon of one neuron projects directly onto the axon of another \cite{Cho2016-br} in such a way that the firing activity arriving from the first axon modulates the synaptic transmission from the second axon toward a third neuron. 
Another important triadic interaction involves the so called tripartite synapse, where astrocytes modulate the synaptic transmission between two neurons by releasing gliotransmitters \cite{Cho2016-br}.

Triadic interactions appearing in complex networks are a general mechanism by which a node can regulate a link between two other nodes. Despite their prevalence in nature -- e.g., in the climate \cite{Boers2019-er} and in ecological networks \cite{Bairey2016,Grilli2017-zv} --triadic interactions have been largely neglected, however, in neuronal network studies. 
As said above, in brain networks, tripartite synapses and axo-axonic connections are prominent examples of triadic interactions that could have profound implications for the development and function of actual neural circuits.  
  
From the theoretical perspective, triadic interactions have become an important research focus in recent years as they have been shown to dramatically change the effective connectivity of a network. 
One prominent example is the recently developed framework of triadic percolation \cite{Sun2023, anapnasnexus, sun2024higher} -- by which nodes in the giant component of a network regulate the state of all links in the network -- which provides a promising model for studying time-dependent network connectivity, such as in brain networks. 
Thus percolation -- determining the giant component -- becomes a dynamic process and the giant component of the network changes over time, resulting in periodic oscillations or even chaos \cite{Sun2023, sun2024higher}. 
In networks with a spatial organization, triadic percolation induced complex spatio-temporal patterns of the effective connectivity \cite{anapnasnexus}: at each time-step, different connectivity patterns emerge with distinct active (connected) and inactive (disconnected) regions. 
These studies provide promising theoretical frameworks for modeling triadic interactions in actual complex networks with time-dependent connectivity such as brain and climate networks.

The theory of triadic percolation can capture global effects of triadic regulations, but not local or activity-driven effects. 
In this context, the triadic percolation framework \cite{Sun2023, anapnasnexus, sun2024higher} also may be extrapolated to neuronal networks by explicitly accounting for neuronal dynamics. In a simple description, neuronal dynamics are described by the membrane potential \cite{burkitt2006review} -- the difference in electric potential between the interior and exterior of the neuron soma -- which changes driven by external and recurrent and currents. 
Once the membrane potential surpasses a threshold, this elicits an action potential or spike of activity \cite{kandelbook,kochbook}.
The activity of simple neuronal networks is influenced by the network connectivity, which drives the recurrent currents. 
However, as oppose to percolation where the giant component captures a global phenomenon, recurrent currents are mostly driven by local connectivity. 
The effect on the emergent activity spatio-temporal patterns of local versus global rules on triadic regulation, as well as of explicitly considering neuronal dynamics, remains thus an open question. 
 
Here we go a step further and consider the case of triadic regulations on a network of leaky integrate-and-fire (LIF) neurons \cite{burkitt2006review, fourcaud2002dynamics}.
We study the emergent spatio-temporal patterns of neuronal activity and network connectivity as the strength of the synaptic interactions is modified. We have found four distinct regimes:
i) a silent phase with very low neuronal activity and neglible connectivity, 
ii) a low-activity phase characterized by spatio-temporal patterns of low neuronal firing rates propagating through the medium, and that could be related with the activity of resting state networks in the brain \cite{DELUCA20061359},
iii) a high-activity phase characterized by spatio-temporal patterns of neurons having high tonic neuronal activity, i.e., high firing rate, and finally 
iv) a phase characterize by the pseudo-periodic switching between a high- and a low-firing-rate states. This latter regime is similar to the up/down state transitions regularly observed in the cortical activity of anesthetized mammals \cite{Torao-Angosto2021-nf, updownbistable}. 
We have characterized in deep the main features of all these emerging phases and related them with the mechanisms responsible for the triadic percolation phenomenon. 

\begin{figure}[t!bh]
    \centering
    \includegraphics[width=0.8\linewidth]{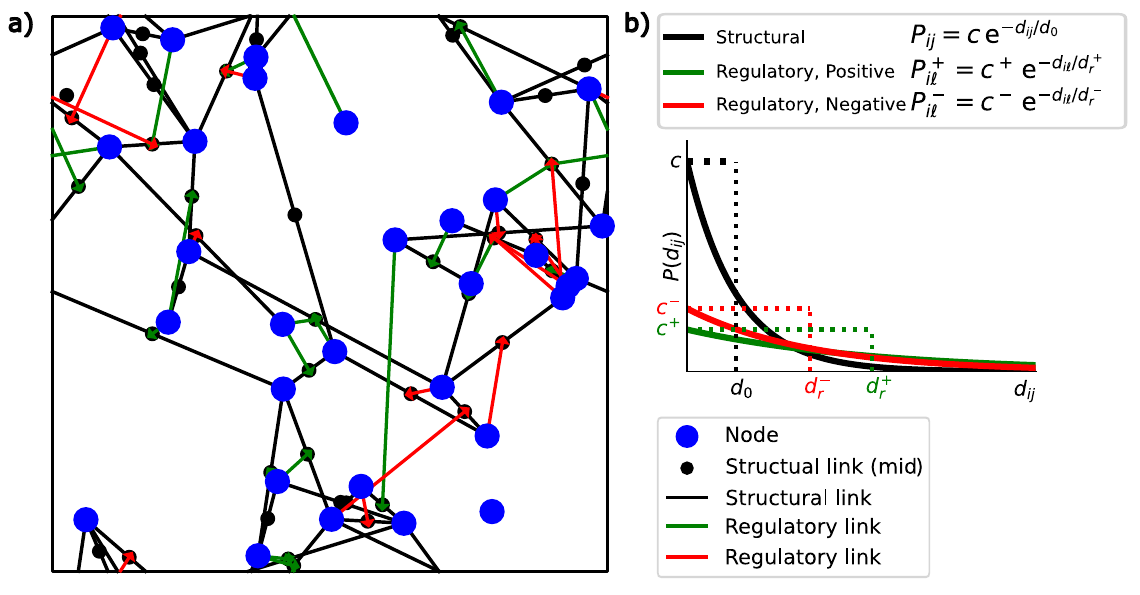}
    \caption{Illustration of the neural network used in the present study: \textbf{a)} Exemplary network with $N=36$, $\rho=100$, $c=1.0$, $d_0=0.08$, $d_r=0.04$, $c^+=c^-=1.0$. Nodes are shown by blue circles, whereas structural links are shown by black lines, and their mid-point by a smaller black circle.
    Structural links are shown by green (positive) and red (negative) arrows from the regulatory node to the regulated link. 
    Periodic boundary conditions are implemented. Links are shown via the shortest direction. \textbf{b)} Connection probabilities for structural and regulatory (both positive and negative) links decay exponentially with distance. In this study we consider the specific case $c^+=c^-$ and $d_r^+=d_r^-=d_r$.
    }
    \label{fig:network_example}
\end{figure}

\section{Models and Methods}

\begin{table}[t!]
    \centering
    \begin{tabular}{|c|c|c|}
     \hline
          \multicolumn{3}{|c|}{Network parameters} \\
      \hline
    \hline
         Parameter & Symbol & Value  \\
         Network size & $N$  & $10^4$ \\
         Scaling factor of structural links& $c$ & $0.4$ \\
         Scaling factor of positive regulatory links& $c^+$ & $0.2$ \\
         Scaling factor of negative regulatory links& $c^-$ & $0.2$ \\
         Typical length of structural links& $d_0$ & $0.25$ \\
         Typical length of positive regulatory links& $d_r^+$ & $0.25$ \\ 
         Typical length of negative regulatory links& $d_r^-$ & $0.25$ \\ 
         Duration of transient period (steps) & $T_0$ & $500$ \\
         Integration time (steps) & $T_\textrm{max}$ & $2500$\\
     \hline
         \multicolumn{3}{|c|}{Neuronal parameters} \\
    \hline
         Parameter & Symbol & Value  \\
    \hline
         External current (constant)& $I_{DC}$ & $0.4$ \\
         External current (noise, standard deviation)& $\sigma_{\textrm{noise}}$ & $0.4$ \\
         Strength of synaptic currents& $A_0$ & Free parameter \\
         Regulation threshold& $\nu_0$ & $0.1$ \\  
         Length of the neuronal time-window & $T$ & $100(*)$ \\
     \hline
    \end{tabular}
    \caption{Typical model parameter values used in most of the analysis reported in the present work. $(*)$: In figure \ref{fig:diagrams_mR} other values of $T$ are tested to dermine its effect on the emergent dynamics.}
    \label{tab:my_label}
\end{table}

The system is defined by an {\em in silico} neural population of $N$ leaky integrate-and-fire (LIF) neurons that interact via synapses. 
Active neurons can regulate synapses either positively or negatively, mimicking e.g. axo-axonic facilitation or inhibition, respectively. 
The modeling scheme builds up on previous theoretical and computational work on triadic percolation \cite{Sun2023,anapnasnexus,sun2024higher}.
In triadic percolation on spatial networks, complex spatio-temporal patterns of connectivity were shown to emerge as a function of the probability that a link is randomly inactive (probability of random damage), which acts as thermal noise. 
For high random damage the connected patterns took the form of local clusters that diffused through the network. 
For low values instead the percolating giant-connected-component is a stripe that crosses the system, and shows a combination of blinking and diffusive temporal dynamics. 
Finally, for intermediate values complex-shaped patterns emerged, which we deemed \emph{octopus} due to their tubular organization, and which blinked in time between two complementary states that spanned the whole system. 
Motivated by these intriguing phenomena, here we investigate the role of triadic interactions in a biologically inspired \emph{in silico} neural medium that shares key features with biological neural networks, to bridge the gap between the theoretical set-up and the biological application. 
Most details of the system are analogous to those in Ref. \cite{anapnasnexus} with the inclusion of one key ingredient, namely the neuronal dynamics,  that drive the triadic interactions in the form of activity-dependent synaptic regulation.

\subsection{Neuronal network}
We consider that neuronal dynamics takes place on a fast time-scale $\tau$, whereas synaptic dynamics occurs on a slower time-scale $t$. 
For simplicity, neuronal dynamics are modelled via the LIF neuron description. That is, the state of each neuron is characterized by its membrane potential $V_i(\tau)$, which evolves in time according to
\begin{equation}
\tau_m\frac{dV_i(\tau)}{d\tau}=-V_i(\tau)+R_m[I_i^{syn}(\tau)+I_{DC}+I_i^{noise}]
\end{equation}
where $\tau_m$ is the membrane time constant, $I_{DC}$ is a constant input current, $I_i^{noise}$ is a random current arriving to each neuron, described by Gaussian white noise with zero mean and standard deviation $\sigma_{noise}$,  and $I_i^{syn}(\tau)=A_0\sum_j \omega_{ij}(t) \epsilon_{ij} s_j(\tau)$ is the synaptic current that neuron $i$ receives from its neighbours. Here $s_i$ is a binary variable state representing the firing state of each neuron being equal to $1$ if neuron $i$ is firing (i.e. $V_i(t)>V_{th}$) and $0$ otherwise. 
The matrices $\epsilon_{ij}$ and $\omega_{ij}(t)$ are binary matrices that encode the structural and functional connectivity, respectively.

The matrix $\epsilon_{ij}$ has elements $1$ and $0$ indicating the presence or lack of a synapse between neurons $i$ and $j$, respectively.
To emulate the spatial dependence of synaptic connections, the elements of $\epsilon_{ij}$ are drawn with probability $P_{ij}$ that depends on the Euclidean distance $d_{ij}$ between them,
\begin{equation}
    P_{ij}=c e^{-d_{ij}/d_0},
\end{equation}
where $0<c<1$ and $d_0$ measures the typical effective distance between two connected nodes. 
An exemplary illustration of the resulting network is shown in Figure \ref{fig:network_example}. 

The state of each synapse, as said above, is subjected to positive (facilitation) and negative (inhibition) regulation. 
The functional state of each synapse is thus encoded by the functional connectivity matrix $\omega_{ij}(t)$, which equals $1$ if a synapse is active, and $0$ otherwise.

\begin{figure}[t!bh]
    \centering
    \includegraphics[width=0.95\textwidth]{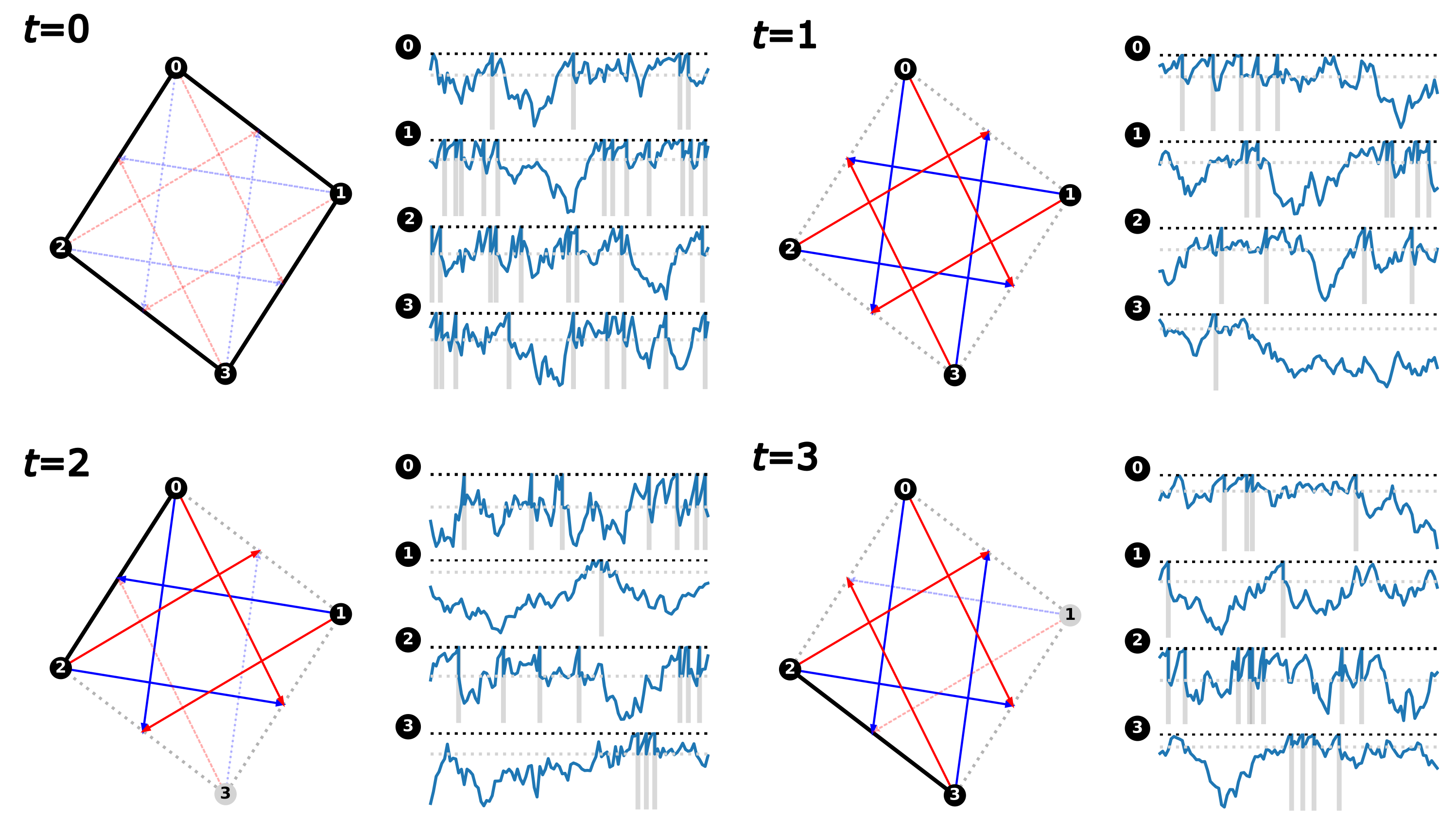}
    \caption{Illustration of the model dynamics. We show here the first four steps of the system dynamics of a regular $4$-node network.
    For $t=0$ we show the network backbone formed by $4$ nodes and $4$ links forming a square. The regulatory backbone comprises $4$ positive and $4$ negative regulatory links placed in a regular manner forming a star as illustrated. 
    The LIF dynamics is initially run on the network backbone for an integration window of length $T$, resulting in the time-series shown on the right-hand side, for each node. Solid-lines show the membrane voltage $V_i(\tau)$, and vertical stripes indicate neuron spikes, $s_i(\tau)=1$. We set a threshold for the activation of the triadic regulations $\nu_0=2/T$. 
    At $t=0$ the firing rate of the $4$ neurons is above the threshold, and all triadic regulations are activated at $t=1$ as shown in the corresponding left-hand-side panel. This results in all structural links having one negative regulation and one positive regulation, and thus are all functionally inactive at $t=1$. The LIF dynamics are run again on this network, resulting on the time-series shown on the right-hand side of panel. In this case only neurons $0$, $1$ and $2$ fire above the threshold and activate their regulatory links. 
    The same process is iterated over time as shown for $t=2$ and $t=3$ panels. }
    \label{fig:dynamics_illustration}
\end{figure}

\subsection{Regulatory networks}
We use the higher-order triadic regulation model recently proposed in \cite{Sun2023,anapnasnexus, sun2024higher} such that an active node (or neuron) can regulate the link (or synapse) between other nodes (see Figure \ref{fig:network_example}). The regulatory mechanism operates in an activity-dependent manner and closes a feedback loop between neuronal and synaptic dynamics. 
A node $i$ can regulate either positively ($+$) or negatively ($-$) (but not both) a link $\ell$. These regulatory links are encoded in the bipartite binary matrices $\epsilon^+_{i\ell}$ and $\epsilon^-_{i\ell}$. 
These are also drawn randomly with probabilities that decay exponentially with the distance between the node and the link, i.e.   
\begin{equation}
    P^+_{i\ell}=c^+e^{-d_{i\ell}/d_r^+}; \quad P^-_{i\ell}=c^-e^{-d_{i\ell}/d_r^-}.
\end{equation}
Here $d_{i\ell}$ is the Euclidean distance between the regulating node $i$ and the regulated link $\ell$, $0<c^+,c^-<1$, and $d_r^+$ ($d_r^-$) is the typical distance at which positive (negative) link regulation can occur. The position of each link is set as the mid-point between the two nodes it connects. 
 
The functional feedback loop is established as follows. 
First, neuronal dynamics takes place on a fixed network $\epsilon_{ij}\omega_{ij}(t)$ for a time-window of length $T$.
Then, neurons with a firing rate $\nu_i(t)\equiv T^{-1} \int_{t-T}^{t} d\tau \; s_i(\tau)$ larger than a threshold $\nu_0$ are deemed to be active and exercise regulation onto the links.  
The state of each synapse is then given by
\begin{equation}\label{eq:triadic}
    \epsilon_\ell(t) = \mathcal{R}_\ell^{+} (1-\mathcal{R}_\ell^{-}),
\end{equation}
where $\mathcal{R}_\ell^{+}$ and $\mathcal{R}_\ell^{-}$ encode the triadic regulations acting onto link $\ell$: $\mathcal{R}_\ell^{+}$ ($\mathcal{R}_\ell^{-}$) equals $1$ if there is at least one active positive (negative) regulation onto link $\ell$, and $0$ otherwise. 
That is, only links that are positively and not negatively regulated are active in the next step $t+1$. 
All model parameter values used in the present work are summarized in Table $1$.
The dynamics is simulated for a macroscopic time $T_\textrm{max}$. To avoid influence of the initial transient, all measures are applied after an initial transient $T_0$.

\section{Results}

The inclusion of triadic regulations on an excitatory LIF neuronal network induces a non-trivial feedback loop between neuronal activity and the emerging functional network topology. Both, positive and negative regulations, are driven by high neuronal activity, and, consequently, activity in one time-step influences the network functional-topology non-trivially in the next, which in turn shapes subsequent neuronal activity.  
In Figure \ref{fig:dynamics_illustration} we show the model dynamics for a simple system. 

In order to emphasize the coupling between neuronal activity and functional topology, we focus here on the regime where neuronal activity is mainly driven by recurrent connections, that is, $A_0\gg I_{DC}$ and $A_0\gg I^{noise}$.
Moreover, and for simplicity, we consider the situation where the number of positive (facilitating) and negative (inhibiting) regulatory links is identically balanced ($c^+=c^-$), and their spatial distribution is equivalent ($d_r^+=d_r^-$). 
We note, however, that due to the asymmetry of the regulatory rule in Eq. \ref{eq:triadic}, this results in a slight net functional bias towards axo-axonic inhibitory regulation. 
Note that in a network of only excitatory neurons, this negative regulation of excitatory synapses helps to prevent the uncontrolled spreading of activity in a seizure-like state. 
A slight bias towards inhibitory regulation is also in line with recent experimental and theoretical observations of brain dynamics, according to which healthy brain dynamics may lie close to a critical point, but in the subcritical regime \cite{fosque2021evidence, buendia2022broad, fagerholm2015cascades}.

At the macroscopic time-scale, we characterize the activity of neurons through the integrated neuronal firing rate at the neuronal level as defined in the previous section, i.e. $\nu_i(t)= \frac{1}{T} \int_{t-T}^t d\tau s_i(\tau)$, and also through its temporal average $\widebar{\nu_i} \equiv  \frac{1}{T_{max}-T_0}\sum_{t=T_0}^{T_{max}} \nu_i(t)$. Here $T_{max}$ is the number of macroscopic steps $t$. We do not consider steps prior to $t=T_0$ to avoid transient effects.
The parameter $T$ is the length of the microscopic integration window (fast time-scale $\tau$).

At the macroscopic level, we measure the network-averaged firing rate $\langle \nu \rangle(t)\equiv \frac{1}{N}\sum_i \nu_i(t)$. 
To illustrate the dynamics at the microscopic time-scale we also measure the network-averaged firing rate at this scale, $\langle \nu \rangle(\tau) \equiv \frac{1}{N} \sum_i s_i(\tau)$. 
In general, note that $\langle \cdot \rangle$ stands for network averages, whereas $\widebar{\cdot}$ stand for time averages. 
Functional connectivity at each network time-step $t$ is characterized by the connected components $C^j(t)$ of size $R^j(t)$. The connected components are ordered from largest (giant connected component, $j=0$) to smallest, so that $R^0$ corresponds to the typical percolation order parameter \cite{Sun2023}.
To analyze the coupling between activity and functional topology, we further measure the activity of the neurons in each connected component, $\langle \nu \rangle^j(t)$.

\subsection{Emergence of functional patterns}

\begin{figure}[tbh!]
    \centering
    \includegraphics[width=0.8\textwidth]{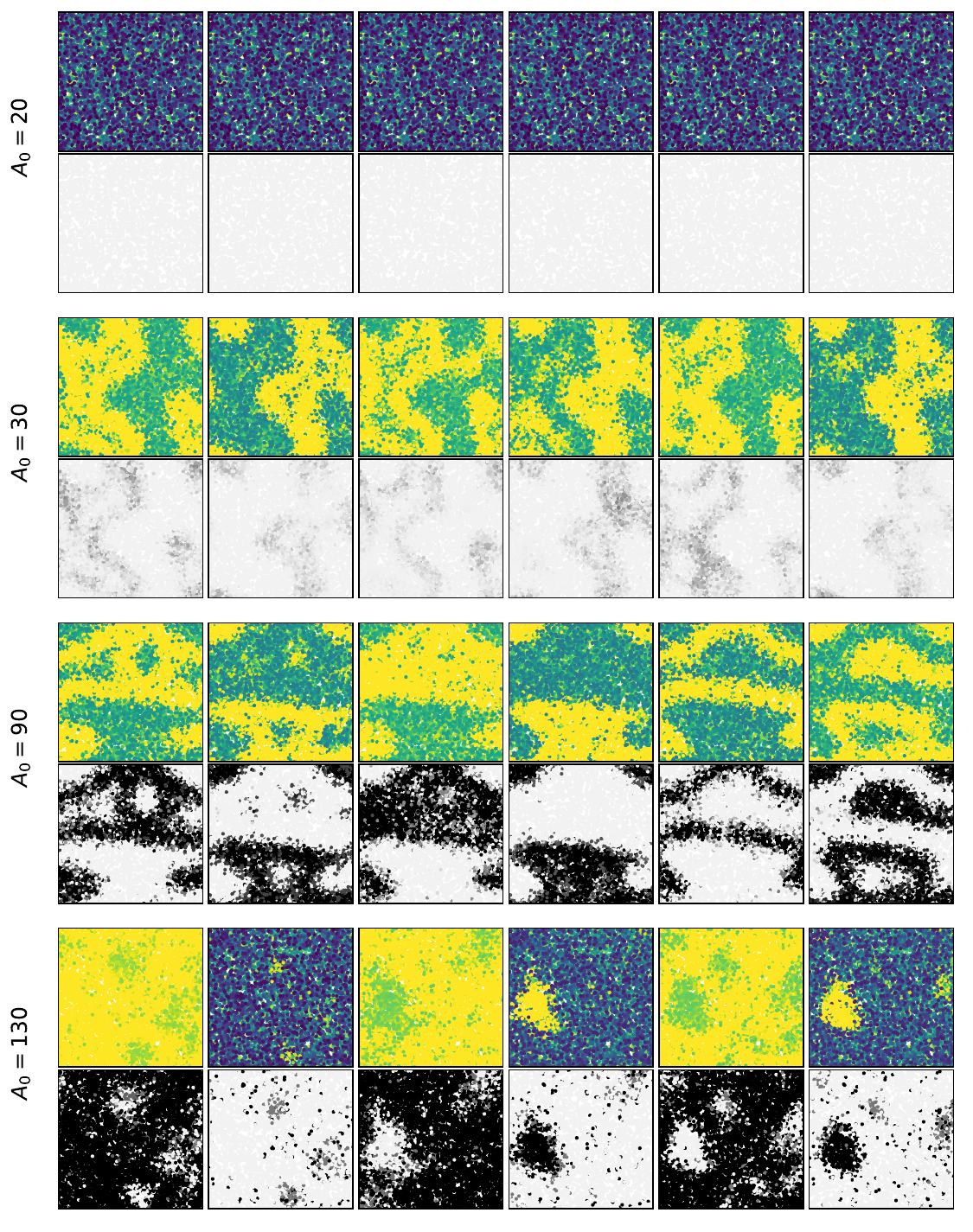}
    \caption{Exemplary activity and functional-connectivity spatio-temporal patterns. 
    For each $A_0$, the top row shows the functional connected components $C^j$. Larger components are shown with light colors (yellow) and smaller ones with dark colors (green and blue). 
    The bottom row shows the neuronal firing rate $\nu_i(t)$ with a grey palette (with black indicating maximum firing rate), 
    We have chosen four representative values of $A_0=20,\ 30,\ 90,\ 130$ (from top to bottom), and six consecutive time-steps (from left to right). }
    \label{fig:patterns}
\end{figure}

\begin{figure}[tbh!]
    \centering
    \includegraphics[width=0.8\textwidth]{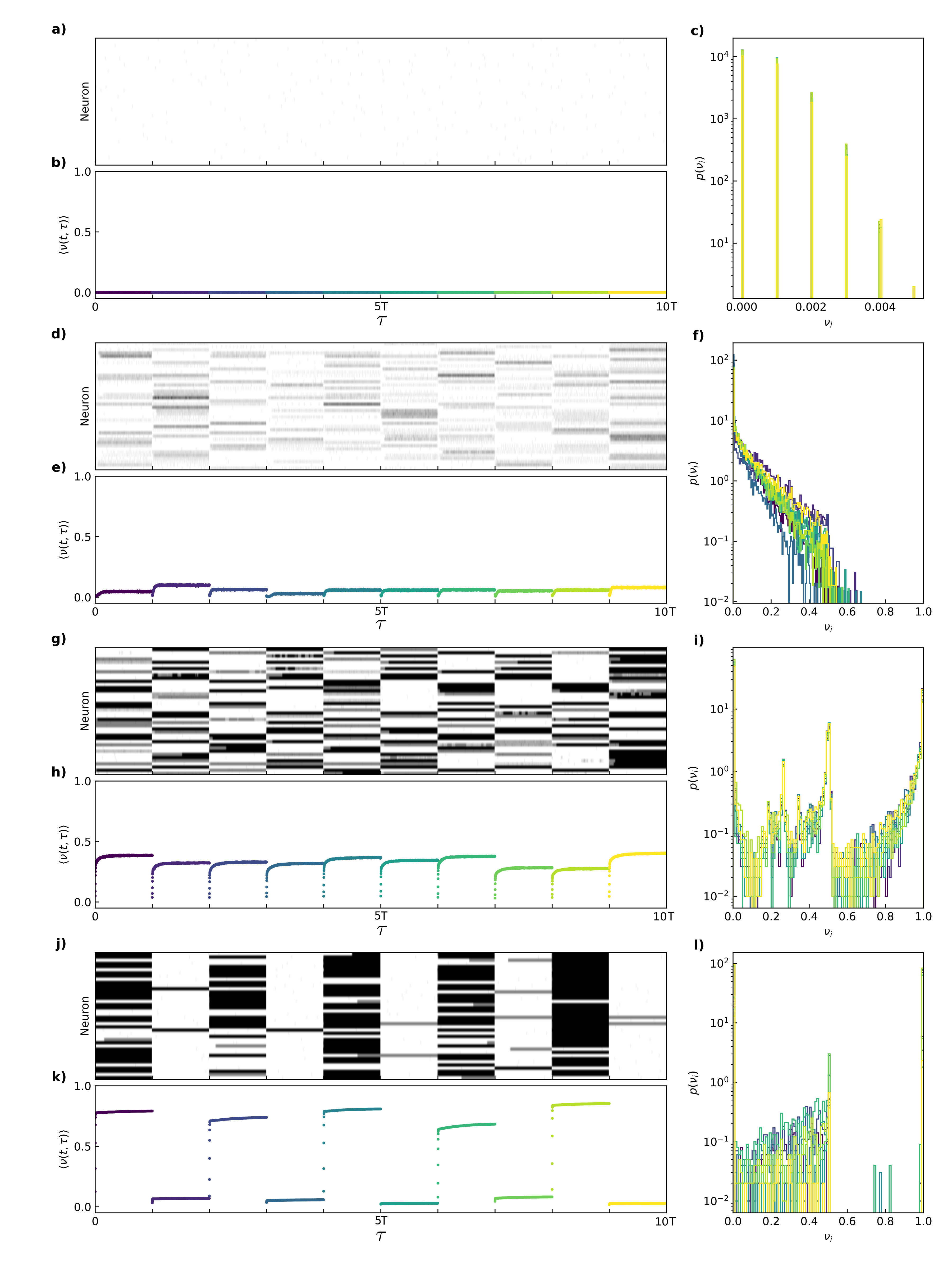}
    \caption{Exemplary activity at the microscopic time-scale $\tau$. Each main row corresponds to a different value of $A_0$ as in figure \ref{fig:patterns}. 
    Panels in the left column show the microscopic time-series of activity via the raster-plots of $40$ randomly-selected neurons (top sub-panels) and via $\langle \nu \rangle(\tau)$ (fraction of neurons that fire at the microscopic time-step $\tau$, for macroscopic window $t$, bottom subpanels). 
    Panels in the right column show 
    the distribution of integrated neuronal firing-rates $p\left( \nu_i(t) \right)$, where the time-average is over the microscopic time-scale $\tau$.
    The color-code indicates the macroscopic time-window, from $t=0$ (dark) to $t=9$ (light). 
    }
    \label{fig:microscopic}
\end{figure}

Activity-driven triadic regulations give rise to complex spatio-temporal patterns of activity and functional network topology.  
Distinct regimes emerge as a function of $A_0$, as shown in Figure \ref{fig:patterns} for four exemplary $A_0$ values. 
For each one, we show $6$ consecutive steps of the functional network and activity states.
Functional connectivity (odd rows) is shown via the connected components $C^j$, by drawing each one in a different color that  indicates $R^j$, with larger components shown in lighter colors.
Neuronal activity is shown by the neuronal firing rate $\nu_i(t)$ (in a grey-scale such that black stands for high activity).  
The microscopic dynamics corresponding to each of the regimes is shown in Figure \ref{fig:microscopic} with
raster plots of activity of a sub-sample of $40$ randomly-selected neurons (left top panels), the instantaneous network firing-rate $ \langle \nu \rangle(\tau)$ (left bottom), and the distribution of neuronal firing rates $\nu_i(t)$ (right). 

Functional connectivity and neuronal activity are strongly influenced by the synaptic strength $A_0$.
For relatively small $A_0$ ($A_0=20$ in Figure \ref{fig:patterns}, panels a, b, c in Figure \ref{fig:microscopic}), the functional network is almost disconnected and there are only small functional connected components. Correspondingly, the firing rate is very small or null. We call this a \emph{silent phase}.
For intermediate $A_0$ ($A_0=30$ in Figure \ref{fig:patterns}, panels d, e, f in Figure \ref{fig:microscopic}), a functional giant component $C^0$ emerges (yellow area in Figure \ref{fig:patterns}) associated with higher neuronal activity. 
Smaller connected components, $C^{1},\ C^{2},$... also emerge, with lower firing rates. The emerging functional topology and neuronal activity patterns present a spatial organization reminiscent of the triadic percolation patterns in spatial networks \cite{anapnasnexus}.
Increasing $A_0$ ($A_0=90$ in Figure \ref{fig:patterns}, panels g, h, i in Figure \ref{fig:microscopic}) leads to the increase in the neuronal firing rate, and does not change the spatial organization of the patterns. We thus distinguish two regimes of spatial patterns: a low-activity and a high-activity one. 
The difference between these two regimes is most apparent in Figure \ref{fig:microscopic}. 
In the high-activity phase, most active neurons are in the tonic regime, with individual firing rates of $0.5$ (switching state) or $1.0$ (tonic state), as shown in panels g and i of Figure \ref{fig:microscopic}.
The distribution of individual firing rates is wide and shows several local maxima (a discussion of this phenomenon can be found in the SI). 
On the contrary, in the low-activity phase, active neurons fire sparsely, and the distribution of individual neuronal firing rates decays exponentially (panels d and f of Figure \ref{fig:microscopic}).

\FloatBarrier

\begin{figure}[tbh!]
    \centering
    \includegraphics[width=0.9\textwidth]{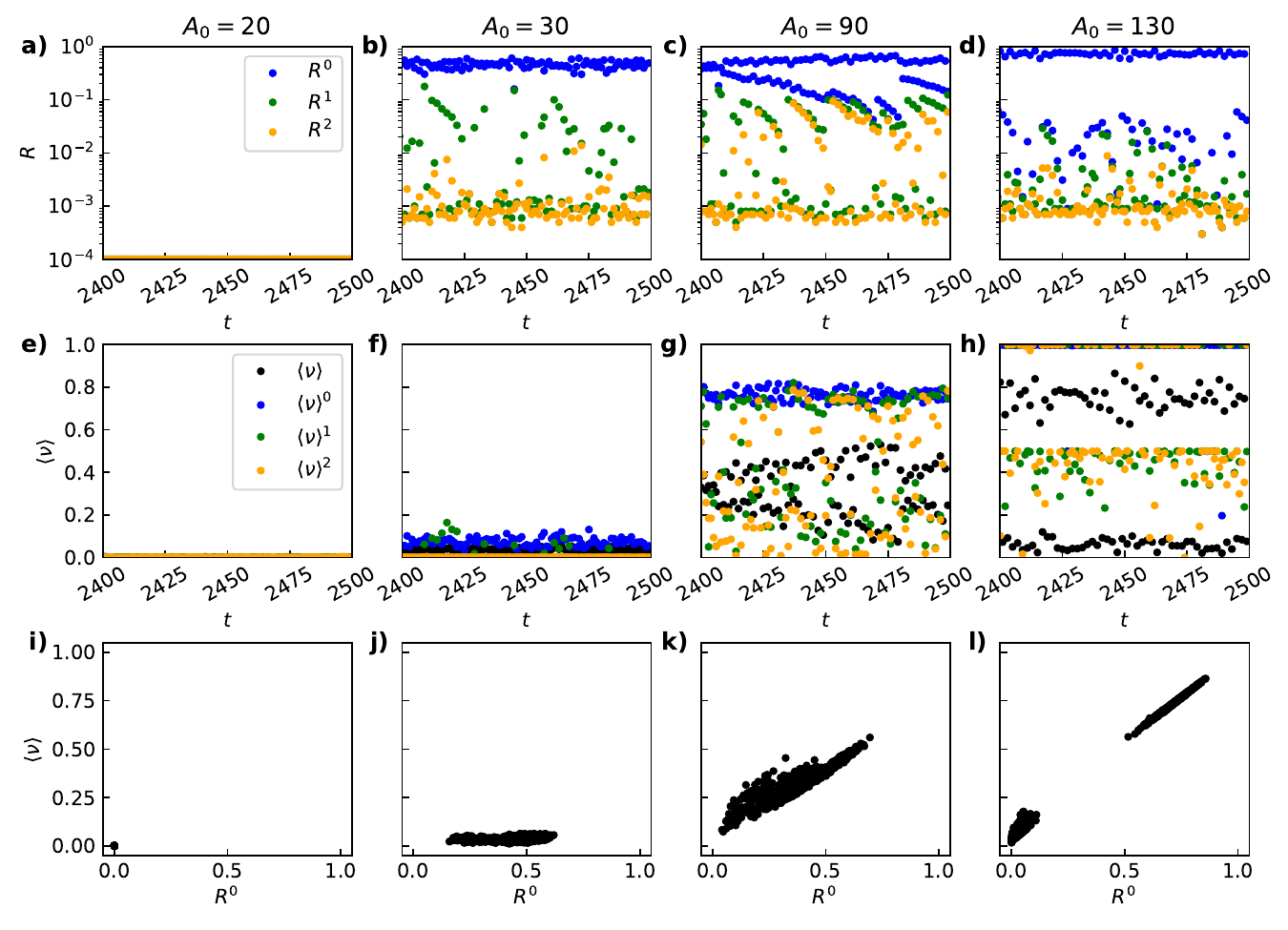}
    \caption{
    Exemplary time-series of the temporal behaviour of the activity and functional-connectivity patterns for the same four representative $A_0$ values as in the previous figures. 
    Panels \textbf{a)} to \textbf{d)} show the size of the three largest connected components (respectively in blue, green and orange for $R^0(t)$, $R^1(t)$ and $R^2(t)$).   
    Panels \textbf{e)} to \textbf{f)} show the network-averaged instantaneous firing rate, $\langle \nu \rangle(t)$, for all neurons (black), and for only neurons in the three largest connected-components (respectively blue, green and orange for $\langle \nu\rangle^0(t)$, $\langle \nu\rangle^1(t)$ and $\langle\nu\rangle^2(t)$). 
    Finally, panels \textbf{i)} to \textbf{l)} show the relation between functional topology and neuronal activity by plotting  $\langle \nu \rangle^0(t)$ versus $R^0(t)$. 
    }
    \label{fig:time_series_mR}
\end{figure}

Finally, for very large values of $A_0$ ($A_0=130$ in Figure \ref{fig:patterns} and panels j, k, and l in Figure \ref{fig:microscopic}), the system \emph{blinks} between two complementary patterns of activity and connectivity: from a state of high network activity and a large giant component to one of low activity and a small giant component.   
The distribution of individual neuronal firing-rates thus presents two main local maxima at $0$ and $1$. There is also a smaller local maximum at $0.5$ preceded by an exponential growth. This phenomenon is also discussed in the Supp. Information (SI). 
This regime closely resembles the short-term blinking regime of triadic percolation: the functional giant component fluctuates from large to small due to the combination of local positive and negative triadic regulation \cite{Sun2023, anapnasnexus}.  Also, the alternation in time of high and low activity periods in this phase resembles the up/down transitions observed in the cortical activity of anesthetized animals \cite{Torao-Angosto2021-nf, updownbistable}. 

\begin{figure}[tbh!]
    \centering
    \includegraphics[width=0.95\textwidth]{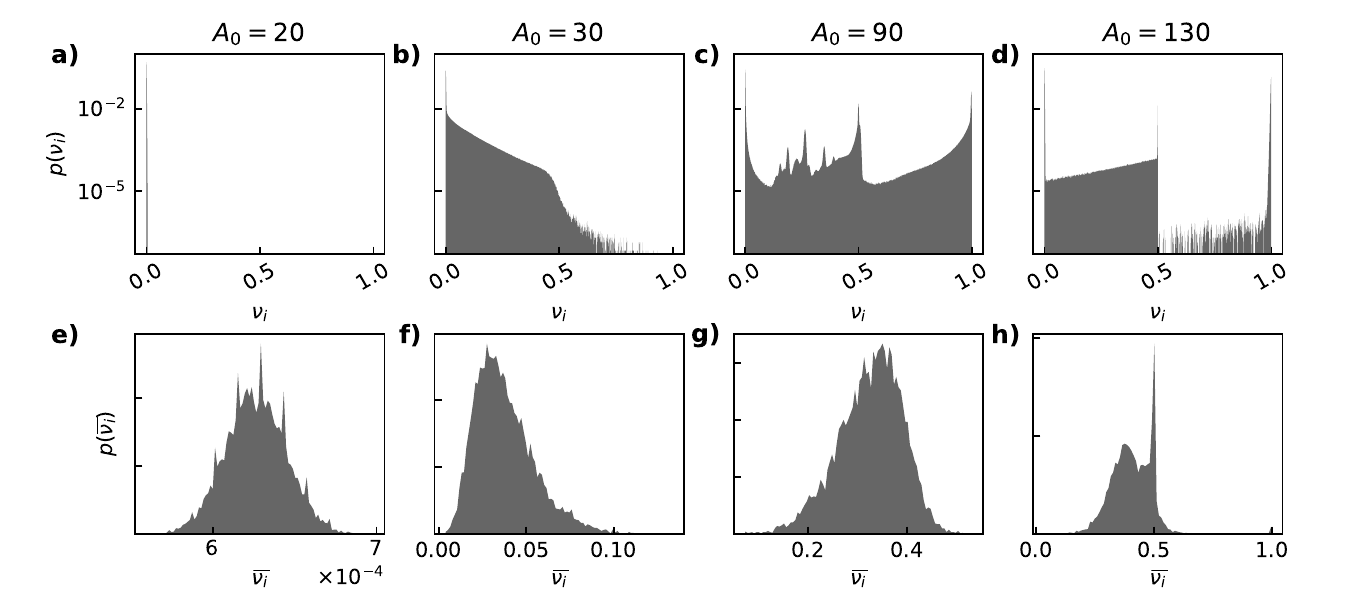}
    \caption{
    Neuronal dynamics features for four representative $A_0$ values as shown in the column-titles. 
    Each row corresponds to a different metric. 
    Panels \textbf{a)} to \textbf{d)} show the distribution of instantaneous neuronal firing rates $\nu_i(t)$ across the neuronal population.
    Panels \textbf{e)} to \textbf{h)} show the distribution of time-averaged neuronal firing rates, $\widebar{\nu_i}$, across the neuronal population. 
    }
    \label{fig:time_series_distributions}
\end{figure}

\subsection{Quantitative description of the activity-topology feedback-loop}

To better quantify the behavior of the system in each dynamical phase, in Figure \ref{fig:time_series_mR} we have described the macroscopic dynamics of neuronal activity and functional topology via the size of connected components $R^j(t)$ (panels a-d), and the mean firing rate of the network, $\langle \nu\rangle(t)$, and those of the three largest connected components $\langle \nu\rangle^j(t)$, $j=0,\ 1,\ 2$ (panels e-h). 
Finally, we show the correlation between $\langle \nu\rangle(t)$ and $R^0$ (panels i-l).

The results in Figure \ref{fig:time_series_mR} confirm our preliminary description of the different regimes. We can draw further insights from this quantitative exploration. 
For the silent phase, no giant connected component emerges and firing rates are low (panel a).
In the \emph{low-activity regime} (panel b), $R^0$ is roughly constant and equal to $0.5$ -- meaning that roughly half of the neurons belong to the giant component at each step. In this regime $C^1$ is small but non-zero, spanning about $1\%$ of neurons, and presents large fluctuations. Finally, $C^2$ remains small, comprising about $0.1\%$ of neurons (which corresponds to around $10$ nodes in this case). 
In the \emph{high-activity regime}, on the contrary, $R^0$ fluctuates between a state dominated by a large $C^0$ to one with several components of smaller size (panel c). 
These regimes correspond to the phase of traveling functional and activity patterns. 
Finally, in the \emph{pseudo-blinking regime} the period-$2$ oscillations of the giant-component become dominant as 
the network oscillates from a state dominated by a giant connected component that spans most nodes ($R^0>0.5$) to an almost disconnected phase ($R^0\to 0$) where many small connected components emerge (panel d). 
We note that this regime does not achieve perfect period-$2$ blinking due to the thermal and quenched disorder that distorts the spatial patterns over long time-scales, as indeed it is also found for triadic percolation on spatial networks \cite{anapnasnexus}.

Neuronal activity follows closely the changes in functional topology. 
In the active phases, we confirm that neurons in the giant component are typically more active in general than neurons in other components (panels f, g, and h). 
In the \emph{pseudo-blinking regime} (panel h) $\langle \nu^0\rangle$ is almost constant and equal to $1$,  indicating that all neurons in $C^0$ are firing tonically. The same is not true for all components, and $\langle \nu \rangle(t)$ fluctuates with high period-$2$ between a high ($\langle \nu \rangle(t)\approx 0.75$) and a lower ($\langle \nu \rangle(t)\approx 0.5$) value. 
Similarly in the \emph{high-activity regime}, we recover the fluctuation between a high-activity and a low-activity state in $\langle \nu\rangle(t)$ as a consequence of the oscillation between a connected (large $C^0$) and fragmented (small $C^0$) network. 

The coupling between neuronal activity and network topology is confirmed by plotting $\langle \nu\rangle(t)$ versus $R^0(t)$. We find a strong linear correlation between the two, particularly in the high-activity and in the blinking regimes: the Pearson correlation coefficients are, respectively, $0.294$, $0.908$ and $0.999$ for the three active phases.

To further understand neuronal dynamics in each regime, in Figure \ref{fig:time_series_distributions} we focus on describing the activity of each neuron, as opposed to the collective description illustrated in Figure \ref{fig:time_series_mR}. 
In order to do so, we deepen the analysis of the distribution of neuronal firing rates of Figure \ref{fig:microscopic}, and display the distribution of individual neuronal firing rates $\nu_i(t)$, as well as the distribution of time-averaged neuronal firing rates, $\widebar{\nu_i}$. 
In this description, the four regimes differ markedly. 
In the \emph{silent regime} most neurons are silent (see panel a), and $\widebar{\nu_i}$ is homogeneously distributed with a very small mean (of magnitude $10^{-4}$) (see panel e).
In the \emph{low-activity regime} $p(\nu_i)$ decays exponentially with $\nu_i$ up to $\nu_i\approx 0.5$, where a cut-off appears (panel b). 
Thus time-averaged firing rate $\widebar{\nu}_i$ is small and its distribution is right-skewed: the majority of the neurons have low $\widebar{\nu}_i$, and few of them display increased firing rates (panel f). 
In the \emph{high-activity regime}, on the contrary, there is a spread-out distribution of $\nu_i$ (panel c), with a large fraction of neurons entering the tonic-firing regime ($\nu_i(t)\to 1$).
Consequently the distribution of $\widebar{\nu_i}$ (panel g) shifts to larger values and is also left-skewed, with most neurons presenting relatively large $\widebar{\nu}_i$. 
Finally, in the \emph{pseudo-blinking regime} the distribution of $\nu_i$ (panel d) is roughly bimodal, with two dominant maxima at $\nu_i = 0.0$ and $1.0$. A third, smaller maximum appears at $0.5$. We recover the exponential increase of $p(\nu_i)$ up to $\nu_i=0.5$, which is discussed in the SI. 
The distribution of $\widebar{\nu_i}$ (panel h) is also bimodal and reflects the pseudo-blinking phenomenon: most neurons have on average a firing rate close to $0.5$ resulting from alternating between constant firing and being silent as they come in and out of the giant component. 

%\clearpage

\begin{figure}[tbh!]
    \centering
    \includegraphics[width=\textwidth]{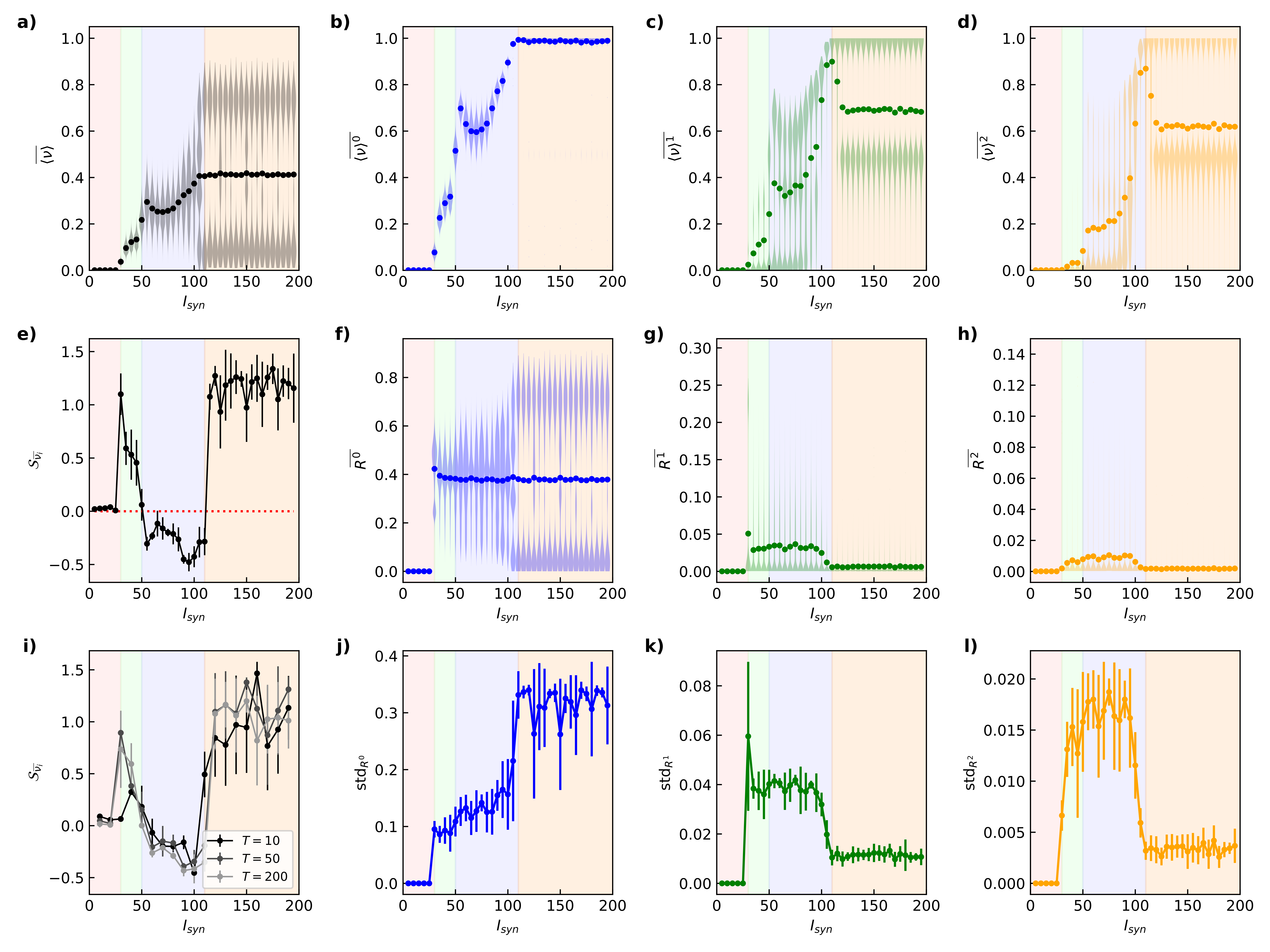}
    \caption{Phase diagram of the topological functional and their corresponding activity patterns as a function of the strength of the synaptic currents, $A_0$. 
    Panels \textbf{a)} to \textbf{d)} show the macroscopic-time average of the network-averaged firing-rate $\langle \nu \rangle(t)$, over all nodes (\textbf{a}), and nodes in the three largest connected-components (\textbf{b} to \textbf{d}), respectively. 
    Panels \textbf{f)} to \textbf{h)} show the size of these three largest connected components. 
    Panels \textbf{j)} to \textbf{l} show the standard deviation of $R^j(t),\ j=0,1,2$, namely $\textrm{std}_{R^j}$. 
    Finally, panels \textbf{e)} and \textbf{i)} show the skewness $\mathcal{S}_{\widebar{\nu_i}}$ of the macroscopic time-averaged neuronal firing rate distributions (as shown in Fig. \ref{fig:time_series_mR}a-d) over all neurons for the standard case (panel \textbf{e},  $T=100$ as considered in all other figures in the main text) and three different values of $T$, spanning both larger and smaller values than the reference one (panel \textbf{i}). 
    In panels \textbf{a} to \textbf{d} and \textbf{f} to \textbf{h}, the data-points stand for the average over realizations ($n=10$) and time-steps (after a transient $T_0=1000$, up the maximum simulation time $T_\textrm{max}=2500$).
    The individual data are represented via the violin plots for each $A_0$, indicating the distribution of values in a summary manner.  
    In the remaining panels, data points stand for the average over realizations, whereas error bars indicate the standard deviation over system realizations. 
    }
    \label{fig:diagrams_mR}
\end{figure}

\subsection{Phase diagram of the activity-driven triadic-LIF model}

The findings in the previous sections are summarized by the phase diagram of the system as a function of the control parameter $A_0$ (Figure \ref{fig:diagrams_mR}). 
First, we show the network-averaged firing rates $\langle \nu \rangle(t)$ and $\langle \nu\rangle^j(t)$ (panels a-d), and the size of the largest connected components $R^j(t)$ (panels f-h). 
The average of these magnitudes over time (at the macroscopic time-scale $t$, indicated by notation $\widebar{\cdot}$, and measured after an initial transient) and system realizations ($n=10$) is shown by data points.
Given that these magnitudes do not converge to a steady fixed-point for large $A_0$, we include violin-plots displaying the distribution of individual data points. This is more representative than a traditional orbit-diagram plot given the large number of individual data-points and the large dispersion of the values.
To quantify this effect, in panels (j-l) we show the standard deviation of $R^j(t)$ over time, namely $\textrm{std}_{R^j}$, for $j=0,\ 1,$ and $2$. 
Finally, to capture the individual neuronal dynamics, we have measured the skewness of the distribution of macroscopic time-averaged firing rates, $\mathcal{S}_{\widebar{\nu_i}}$, which is shown in panel e. 

The combined phase diagrams clearly confirm the emergence of the four main regimes described above. 
The \emph{silent regime} is characterized by null values of all variables.  
There is then a transition into the \emph{low-activity regime} which is continuous on the firing rates but discontinuous on $R$, $\textrm{std}_R$ and $\mathcal{S}_{\widebar{\nu_i}}$. 
The low-activity regime is characterized by low but increasing $\langle \nu \rangle(t)$ and intermediate values of $R^0(t)$. Both  $\langle \nu \rangle(t)$ and $R^0(t)$ are uniformly distributed over time, with unimodal distributions in the violin-plots (and small dispersion in the case of $\langle \nu \rangle(t)$) and a well-defined mean value. This regime further shows small but increasing standard deviation $std_R^j$, and positive $\mathcal{S}_{\widebar{\nu_i}}$. $\widebar{R^1}$ and $\widebar{R^2}$ are small but non-zero in this regime, and present relatively large temporal fluctuations as given by $\textrm{std}_{R^1}$ and $\textrm{std}_{R^2}$.
Both $\widebar{R^1}$ and $\textrm{std}_{R^1}$ peak at the transition point. 
Notably, $\widebar{R^0}$ is approximately constant and equal to $0.5$ in the three active regimes. 

The transition into the \emph{high-activity regime} is not observed in $R^j(t)$ excepted for a subtle widening of the distributions and continuous increase of $\textrm{std}_{R^j}$, but takes place as a discontinuous transition followed by a maximum of
$\widebar{\langle \nu \rangle}$ and $\widebar{\langle \nu \rangle}^j$. It is also shown in $\mathcal{S}_{\widebar{\nu_i}}$ with a transition to negative skewness.
Finally the transition into the \emph{pseudo-blinking regime} is marked by a peak in $\widebar{\langle \nu \rangle}^1$ and $\widebar{\langle \nu \rangle}^2$ and a continuous transition of $\widebar{\langle \nu \rangle}^0$ and  $\widebar{\langle \nu \rangle}$ to a constant value (approximately $0.4$). 
All network-averaged firing rates converge on average to a constant value (which is $1$ for the giant component, indicating constant firing, and smaller for the rest). However, $\langle \nu \rangle(t)$ and $\langle \nu \rangle^2(t)$ continue to fluctuate in time and show bimodal distributions. 
Notably however, in this regime $C^1$ and $C^2$ become negligible again, as the system alternates between the disconnected (no giant component) to the connected (giant component spans the whole or almost the whole network) states. 
Finally, in this regime the skewness of the neuronal firing rates $\mathcal{S}_{\widebar{\nu_i}}$ transitions back to positive values.   

To finalize the analyses, in panel i we report $\mathcal{S}_{\widebar{\nu_i}}(A_0)$, which is the magnitude that best captures the different regimes, for different lengths $T$ of the microscopic time-window, controlling the relative time-scales of the neuronal and network dynamics. As it can be seen, all main aspects of the phase diagram hold when $T$ is modified.

\section{Conclusions}

Triadic interactions give rise to complex spatio-temporal patterns of  connectivity, as was recently shown for the case of triadic percolation \cite{Sun2023, anapnasnexus, sun2024higher}. 
Here, we have taken a step further and explicitly considered the role of triadic interactions in the activity and connectivity of {\em in silico} neural medium -- more precisely a complex network of LIF neurons. Such setup drives a dynamic feedback loop between neuronal activity and functional synaptic connectivity such that neuronal activity shapes the topology, and in turn the resulting topology determines the subsequent activity \cite{fricker2009adaptive, millan2018concurrence}. 
Triadic interactions provide a new means for synaptic plasticity in which, crucially, a neuron may modulate a synapse between another two neurons. 
In actual neuronal populations, for example in the brain, such role is undertaken by the so-called tripartite synapse and axo-axonic connections. The setup considered here provides an abstract model for axo-axonic connections but can be easily extended to account for medium including tripartite synapses. These provide unique functional properties to neural circuits by allowing for a short circuit that bypasses somato and axo-dendritic processing \cite{cover2021axo}.
Axo-axonic synapses also provide precise modulation with a high spatial precision of the specific synapses \cite{cover2021axo}.

Through extensive simulations in our {\em in silico} neuronal medium, we have identified the emergence of four distinct activity regimes or dynamical phases depending on the strength of the synaptic interactions which is related to maximal synaptic conductances: \textbf{a)} first a \emph{silent phase} of low or null neuronal activity and network connectivity; \textbf{b)} second a \emph{low-activity phase} of low activity and sparse connectivity that change in time displaying complex spatio-temporal patterns;
\textbf{c)} third a \emph{high-activity phase} in which the spatio-temporal patterns of activity and connectivity remain, but neural firing rates are high and close to the saturation level; and finally
\textbf{d)} a \emph{pseudo-blinking phase} in which the patterns of high activity and connectivity alternate between a large (spanning almost the full network) and a small cluster. The resulting network firing rate closely mirrors the up/down state transitions in the cortical activity of anesthetized mammals. Interestingly, in this fourth phase there is a strong reorganization of the functional connectivity between up and down states which is compatible with functional connectivity variations observed during wake-sleep transitions in humans \cite{Nguyen2018, Daneault2021}, particularly in young individuals \cite{Daneault2021}, and which origin is similar to the activity and functional connectivity feed-back loop mechanism introduced in the present work \cite{WATANABE20141}.  
 
Three aspects are crucial for the emergence of the dynamical phases described here.
Firstly, a balance between positive and negative regulation. 
As discussed in our previous studies on triadic percolation \cite{anapnasnexus}, a strong deviation from the balanced triadic network destroys the complex spatio-temporal dynamics. 
This finding is in agreement with the general observation of excitation-inhibition (E-I) balance in neuronal and brain networks  \cite{munoz2018colloquium, lynn2021broken}. 
This refers to the delicate equilibrium between excitatory and inhibitory neuronal activity in the brain that characterizes healthy brain activity \cite{munoz2018colloquium}. 
Epileptic seizures, for instance, are characterized by increased cellular excitability which is, at least in part, caused by increased excitation that break E-I balance.  
In our scenario, all neurons are excitatory, and this balance emerges naturally due to the triadic interactions that regulate the state of the synapses:
structurally positive and negative regulations are balanced, whereas functionally high neuronal activity leads to more regulation, both positive and negative. 
Negative regulation (similar to axo-axonic inhibition) thus provides here a stabilizing mechanism that prevents excessive firing.
Our study thus provides a new theoretical background for the study of the effects of plasticity on E-I balance, which has recently emerged as a versatile and efficient ingredient for neural network development \cite{asopa2023computational}. 

The second crucial aspect is the spatial property of the structural and regulatory networks.  
This creates a network with large diameter (i.e. the networks are not small-world) and enables distinct functional regions to emerge on the network \cite{millan2018complex, moretti2013griffiths}. Ultimately, this  allows for the complex spatio-temporal dynamics \cite{anapnasnexus}.
And thirdly, on the local nature of the triadic regulation rule. 
In the case of triadic percolation, even though links are regulated locally, active nodes are chosen with a global rule: belonging to the giant component. In our current case, on the contrary, active nodes are chosen according to a local rule: high node activity. On a network of excitatory LIF neurons, both conditions are related but differ, since local connectivity has the major effect on neuronal firing rates. 
The consequences of this small difference are far-reaching. 
In the triadic percolation model, global stripes always emerged in a spatial network with balanced triadic regulations in the case with no random link deactivation (akin to the high $A_0$ scenario in this study), and small local clusters in the case of high random deactivation (similar to small $A_0$ here), with spatially complex patterns appearing in between. In our case, on the contrary, the transition induced by $A_0$ directly goes from the silent phase to spatially-complex patterns, without the emergence of local clusters. Similarly, stripes, which describe rigid global organization, do not emerge here. 
Overall, by allowing for a local and more biologically plausible activation role, the emergent spatio-temporal patterns of activity and functional connectivity also approximate better the spatio-temporal patterns of healthy brain activity.

In conclusion, the study shows that activity-driven triadic regulation induces complex spatio-temporal patterns of neuronal activity and functional network connectivity. This work bridges the gap between the theoretical framework of triadic percolation and real-world neuronal dynamics, offering insights into how higher-order interactions contribute to neural network function and potentially opening new avenues for modeling brain-like systems.

\section{Acknowledgements}

This work has been supported by Grant No. PID2023-149174NB-I00 financed by the Spanish Ministry and Agencia Estatal de Investigación MICIU/AEI/10.13039/501100011033 and ERDF funds (European Union) (to A.P.M. and J.J.T.).
A.P.M. acknowledges financial support by the ``Ram\'on y Cajal'' program of the Spanish Ministry of Science and Innovation (grant RYC2021-031241-I). H.S. is supported by the Wallenberg Initiative on Networks and Quantum Information (WINQ).
A.P.M. and H.S. would like to thank the Isaac Newton Institute for Mathematical Sciences, Cambridge, for support and hospitality during the programme Hypergraphs: Theory and Applications, where work on this paper was undertaken.
The authors are also thankful to Ginestra Bianconi for insightful feedback on this work.

\bibliographystyle{plain} 
\bibliography{biblio}

\clearpage

\section*{Supplementary Information}

\subsection*{On the scaling of $p(\nu_i)$}

Here we discuss the distribution of neuronal firing rates $p(\nu_i)$ in the pseudo-blinking regime (panel d in figure \ref{fig:time_series_distributions}). 
In Supp. Figure \ref{fig:sup_fig_corr} (panel a) we show $p(\nu_i)$ again for the pseudo-blinking regime ($A_0=130$) but for a longer time-series ($t_{max}=5000$ instead of $2500$ as in the main text).
This shows two main maxima at $\nu_i=0.0$ and $1.0$, which correspond to neurons that fire constantly or none at all in a given integration window. There is a secondary maximum at $\nu_i=0.5$, indicating microscopic switching between the active ($s_i=1$) and inactive ($s_i=0$) states. Between $\nu_i=0.0$ and $\nu_i=0.5$ there is an exponential increase of $p(\nu_i)$. 

To understand this shape of $p(\nu_i)$, we first note that there is a non-linear correlation between the functional degree of a node, $k_i(t)$, and its firing rate $\nu_i(t)$, as shown in Supp. Figure \ref{fig:sup_fig_corr}, panel b (which corresponds to the same simulation as panel a). 
Neurons with $k_i=0$ are silent, whereas neurons with $k_i>5$ fire constantly. This leads to the two main maxima of $p(\nu_i)$.
This is because $A_0$ is set much larger than $I_{DC}$ and $\sigma_\textrm{noise}$ (parameters setting the external current arriving at each neuron). In particular, we note that
i) the external current is such that each neuron has a very low probability of firing if no pre-synaptic spike is received, and ii) in the pseudo-blinking regime one pre-synaptic spike is enough to elicit a post-synaptic spike. 

We consider for instance a pair of two connected neurons, $i=0,1$. 
In the absence of noise the silent state $s_0=s_1=0$ is stable, and so is the active $s_0=s_1=1$ state, since the firing of each neuron at time-step $\tau$ causes the other one to fire at time-step $\tau+1$. Similarly, the period-two cycle where the neurons are in complementary states, that is, $s_0(\tau)=1$, $s_1(\tau)=0$, and $s_0(\tau+1)=0$, $s_1(\tau+1)=1$ is also stable. This explains the three maxima of the probability distribution. 
The exponential increase up to $\nu_i=0.5$ is a consequence of the transient microscopic dynamics that follow each topology update at the macroscopic time-scale. 
In the pseudo-blinking regime the new (at time $t+1$) giant-connected-component does not overlap with the previous one (at time $t$), and therefore many neurons in the giant connected-component do not receive any initial input from their neighbors. 
The neurons do receive a weak noisy external input, which can elicit a spike. Once a spike is elicited, the system falls into the stable two-cycle. The probability that a spike occurs due to the external noise at a given time $t_{1st}$ follows a Poisson distribution \cite{sirovich2015new, alili2005representations, sacerdote2013stochastic}, as we show numerically in Supp. Figure \ref{fig:sup_fig_corr} (panel c).
The observed neuronal firing $\nu_i$ is thus $(1-t_{1st})/2$, and the probability increases exponentially, as observed.

Finally, we note that the two neurons may spike spontaneously at the same time, in which case the system falls into the stable fixed-point $s_0=s_1=1$. There is thus a secondary exponential increase of $p(\nu_i)$ from $\nu_i=0.5$ to $1.0$ which is only hinted in figure \ref{fig:time_series_distributions} (panel d), but can be clearly observed in panel a of Supp. Figure \ref{fig:sup_fig_corr}. 

\renewcommand{\figurename}{Supp. Figure}
\setcounter{figure}{0}
\begin{figure}
    \centering
    \includegraphics[width=\textwidth]{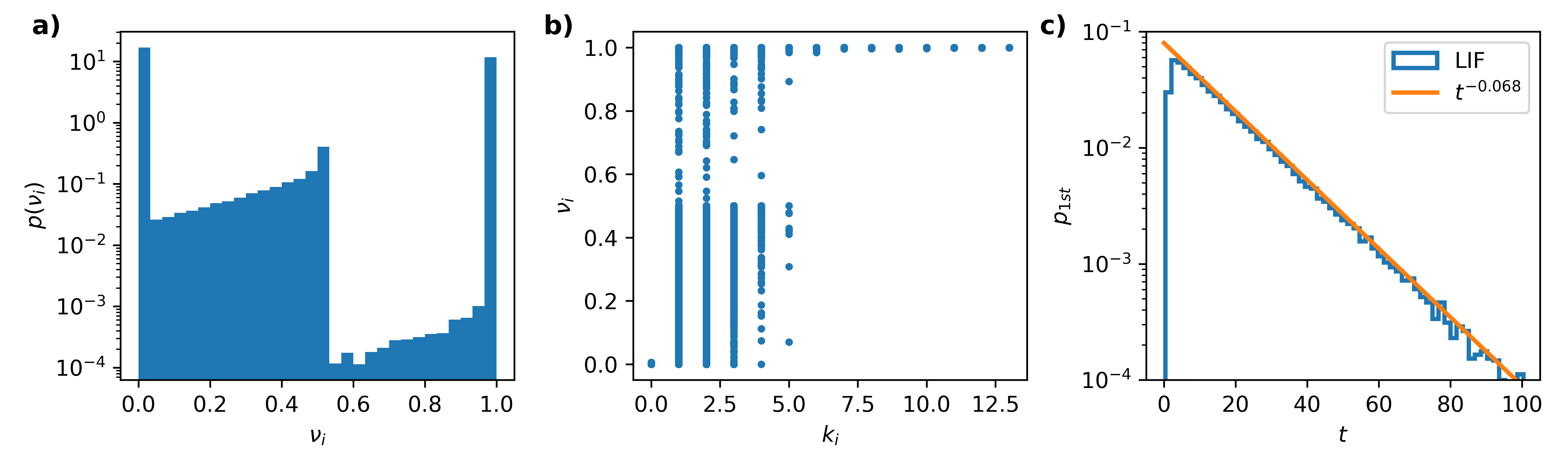}
    \caption{Scaling of the neuronal firing rates in the pseudo-blinking regime. \textbf{a)} Probability distribution of $\nu_i$ on a simulation with $A_0=130$, $N=10^4$ and $T_{max}=5000$, and all other parameters as in the main text. 
    \textbf{b)} Relation between $\nu_i(t)$ and the neuron's functional degree $k_i(t)$ (quantifying the number of active synapses at neuron $i$ at time-step $t$), for the same simulation as in panel a. 
    \textbf{c)} Probability distribution of the first-spike times $t_{1st}$ of a neuron receiving an external input $I_{DC}+I^r_i$ as in our simulations. The orange line indicates an exponential fit with exponent $-0.068$. }
    \label{fig:sup_fig_corr}
\end{figure}

\end{document}